\documentclass{llncs}
 
\usepackage{microtype}
\usepackage{amssymb}
\usepackage{amsmath}
\usepackage{latexsym}

\title{Keeping a Crowd Safe: On the Complexity of Parameterized Verification \\
(Corrected version)\thanks{A former version of this paper was published in the Proceedings of STACS 2014
\cite{DBLP:conf/stacs/Esparza14}. Section \ref{subsec:rv} and Section \ref{subsec:gswol}
contain two mistakes, which are corrected here.}}

\author{Javier Esparza}
\institute{Faculty of Computer Science, Technical University of Munich\\
  Boltzmannstr. 3, 85748 Garching, Germany\\
  \texttt{esparza@in.tum.de}}

\def\by#1{\mathop{{\hbox{\setbox0=\hbox{$\scriptstyle{#1\quad}$}{$\buildrel{\>#1\>}\over{\hbox to \wd0{\rightarrowfill}}$}}}}}

\def\Nat{{\mathbb{N}}}

\begin{document}

\maketitle

\begin{abstract}
We survey some results on the automatic verification of parameterized programs 
without identities. These are systems composed of arbitrarily many components, 
all of them running exactly 
the same finite-state program. We discuss the complexity of deciding that no 
component reaches an unsafe state. 
The note is addressed at theoretical computer scientists in general. 
\end{abstract}

\section{Introduction}

Parameterized programs (where ``program'' is used here in a wide sense) consist of arbitrarily many instantiations of the same piece of code. We call each of these instantiations a {\em process}, and the set of processes a {\em crowd}. Examples include many classical distributed algorithms (for mutual exclusion, leader election, distributed termination, and other problems), families of hardware circuits (for instance, a family of carry-look-ahead adders, one for each bitsize), cache-coherence protocols, telecommunication protocols, replicated multithreaded programs, algorithms for ad-hoc and vehicular networks, crowdsourcing systems, swarm intelligence systems, and biological systems at the molecular level. 

If automatic verification is not your field of expertise, then you may think it awkward to study the complexity of verification problems for parameterized programs. Since Rice's theorem shows that any non-trivial question on the behavior of one single while-program is undecidable, is there any more to say? Actually, yes. Rice's theorem refers to while-programs acting on variables over an infinite domain (typically the integers). Since the primary task of distributed algorithms or cache-coherence protocols is not to compute a function, but rather solve a coordination problem, they typically use only boolean variables as semaphores, or variables ranging between 0 and the number of processes. So for each number $N$, the set of reachable configurations of the crowd with $N$ processes is finite, and most verification questions can be decided by means of an exhaustive search of the configuration space. 

However, this brute force technique can only show correctness for a finite number of values of $N$. This is not what we usually understand under ``proving a parameterized program correct'', we mean proving that the property holds  for {\em all} values of $N$. In other words, the task consists of proving that each member of an infinite family of systems, each of them having a finite state space, satisfies a given property. Are questions of this kind always undecidable?

In the way we have formulated the problem, the answer is still negative: yes, all non-trivial problems are still undecidable. Let us sketch a proof for a simple reachability problem. Given a Turing machine $M$ and an input $x$, we can easily construct a little finite-state program that simulates a tape cell. The program has a boolean variable indicating whether the head is on the cell or not, a variable storing the current tape symbol, and a third variable storing the current control state when the head is on the cell (if the head is not on the cell the value of this variable is irrelevant). A process running the program communicates with its left and right neighbors by message passing. If $M$ accepts $x$, then it does so using a finite number $N$ of tape cells. Therefore, the crowd containing $N$ processes eventually reaches a configuration in which the value of the control-state variable of a process is a final state of $M$. On the contrary, if $M$ does not accept $x$, then
  no crowd, however large, ever reaches such a configuration. So the reachability problem for parameterized programs is undecidable. 

But this proof sketch contains the sentence ``the program communicates with its left and right neighbors''.
How is this achieved? A communication structure where processes are organized in an array (like in our simulation of $M$), in a ring, a tree, or some other shape is achieved by giving processes an {\em identity}, typically a number in the range $[1..N]$. This identifier appears as a parameter $i$ in the code, and so it is not the case that all processes execute {\em exactly} the same code, but the code where the parameter is instantiated with the process identity. For instance, the instruction ``if you're not the last process in the array, then send the content of variable $x$ to your right neighbor'' is encoded as ``if $i < N$, then send the content of variable $x$  
to process $i+1$''. (Observe that, since $N$ also appears in the code, the processes also know how many they are.)

There are applications where processes have no identities and do not know---or do not care about---how many they are: for 
instance, in natural computing processes may be molecules swimming in a solution. In others applications identities are not needed. A typical example are cache-coherence protocols, whose purpose is to guarantee the consistency of all cache lines containing copies of a memory cell. The protocol should guarantee that if a processor updates a variable in its cache, the other processors mark their cached values as no longer valid. Since the situation is completely symmetric, and processors are connected  by a bus implementing a broadcast communication primitive, identities are not needed. The same holds for many multithreaded programs where one only cares about, say, the number of threads that are still active. Finally, there is an increasing number of applications where identities are considered {\em harmful}. For instance, in vehicular networks cars may communicate with each other to interchange information about traffic jams. Since cars must necessarily communicate their positions, identities might allow one to track individual cars. Applications involving secret voting are another example. 

These considerations lead us to our problem, which can be informally,
but suggestively, formulated as follows: 

\begin{quote}
{\bf What is the complexity of checking that a (finite, but arbitrarily large) anonymous crowd will
stay safe?}
\end{quote}
 
Formally, the input to the problem is a finite automaton $A$, the {\em template}, representing the finite-state code to be executed by each process, and a state $q_u$ of $A$, the {\em unsafe state}, modelling some kind of error or undesirable situation. The transitions of $A$ correspond either to internal moves or to communications with the rest of the system. The question to be answered is whether there exists a number $N$ such that some execution of the system composed by $N$ identical copies of $A$ reaches a configuration in which at least one of the processes is in the unsafe state $q_u$.  We say that such configurations {\em cover} $q_u$, and so the problem is called the {\em coverability} problem. 

The complexity of the coverability problem crucially depends on the power 
of the communication mechanism between processes. So first we must analyze 
these mechanisms in some detail. This is done in Section \ref{sec:paradigms}. Section \ref{sec:power} presents the complexity results. Finally, Section \ref{sec:extensions} briefly describes some additional work in which the template $A$ is allowed to have more computational power than that of a finite automaton.

\section{How Crowds Communicate}
\label{sec:paradigms}

The two main communication paradigms are message-passing 
(typical of communication protocols and distributed systems 
where processes reside in different machines) and communication through global variables (typical of multithreaded programs). Within each paradigm there 
is a number of mechanisms. We informally describe the syntax and operational 
semantics of the template $A$ for the four mechanisms most commonly found in the literature. In particular, we give the syntax 
of the transition labels of $A$, and describe how a communication takes place. We assume a finite set $V$ of values which can be communicated.

\paragraph{Broadcast communication.}  Transition labels: $v!!$, $v??$.\\
We assume that for every state $q$ and every value $v$ the template $A$ has one transition $q \by{\mathbf{v??}} q'$ for some state $q'$ (which may be equal to $q$). In a communication step of the system {\it all processes make a move}. Exactly one of the processes takes a transition labelled by $\mathbf{v!!}$, with the intended meaning that this process broadcasts the value $v$ to all others; simultaneously, all other processes take $v??$-transitions, 
depending on their current states.

\paragraph{Rendez-vous communication. } Transition labels: $v!$, $v?$. \\ In a 
communication step of the system, exactly two processes make a move: a process takes a transition
labelled by $\mathbf{v!}$, and, simultaneously, another process takes a transition labeled by $\mathbf{v?}$. The intended meaning is that the first process sends the value $v$ to the second process.

\paragraph{Communication by global store.}  Transition labels: $w(v)$, $r(v)$. \\
In this paradigm we assume that all processes in the crowd communicate with a 
global store. At every point in time the store contains an element of $V$. 
In a communication step, exactly one process makes a move.  The process either takes a transition labeled by $w(v)$, which writes $v$ into the store, or, if 
the current value of the store is $v$, it takes a transition labeled by $r(v)$, meaning that it reads the value $v$ from the store.

\paragraph{Communication by global store with locking. } Transition labels: ${\it lock}$, ${\it unlock}$, $w(v)$, $r(v)$. \\
A process must first obtain a lock on the store before being able to write or read. The processes keeps the lock until it releases it by means of a
transition labeled by ${\it unlock}$. While in possession of the lock, 
the process is the only one that can perform reads and writes.\\

We shall see that the complexity of the coverability problem depends 
on two parameters of the communication mechanism:

\paragraph{(1) Who listens when a process speaks ?} When a process sends a message,
different mechanisms provide different guarantees on who will receive it,
and we can classify them accordingly:
\begin{itemize}
\item[--] {\bf Everyone listens.} This is the case of broadcast communication.
\item[--]{\bf At least someone listens.} This is obviously the case of rendez-vous, 
but also of global store with locking. Indeed, we can easily use a 
global store with locking to simulate rendez-vous communication. 
The store initially contains a special value, say $f$, standing for ``store is free''. 
A process wishing to communicate a value $v$ acquires the lock, reads the content of the store, and, if its value is $f$, changes it to $v$ and releases the lock. If the value is not $f$, it just releases the lock. A process wishing to receive a value acquires the lock and reads the store: if its value is $f$, the process just releases the lock; 
otherwise, it copies the value into its local state and releases the lock. This 
guarantees that the value will be preserved until someone reads it, and, under
a suitable fairness assumption, that it will eventually read.

However, neither rendez-vous communication nor global store with locking 
can implement broadcast. Intuitively, in these paradigms there is no way to
detect that a process does not react to any message.

\item[--] {\bf No guarantee.} This is the case of a global store without locking. A value written by a process 
can be overwritten by another process before anyone reads it. Notice that we can no longer implement rendez-vous using the trick above. Since the store cannot be locked, two processes $P_1$ and $P_2$ wishing to write values $v_1, v_2$ may both read the value $f$ and 
proceed to write. If $P_1$ writes immediately before $P_2$, then the value $v_1$ is not read by anyone.
\end{itemize}

\paragraph{(2) Can the crowd produce a leader?} Loosely speaking, this is the question whether a perfectly symmetric crowd in which initially
all processes are in the same state can be forced to split into a 
distinguished process which stays within a special subset of states of 
the template, and an arbitrarily large crowd that
stays within another subset. More precisely (but still a bit informally) the question is the following. Is there a template $A$ with two distinguished states $q_1, q_2$ and all processes initially in $q_1$, such that some reachable configuration has one process in $q_2$, and no reachable configuration has more than one process on $q_2$?

Broadcast communication and communication through global store with locking can 
both easily produce a leader. In the case of broadcast communication, the 
template with transitions $q_1 \by{a!!} q_2$ and $q_1 \by{a??} q_3$ already does the trick. The process broadcasting the message 
moves to $q_2$ and, since all other processes {\em must} listen, they all move to $q_3$. 
In the case of global store, we choose a template in which  all processes initially compete for the lock; the process that acquires it changes 
the value of the store to ``we have a leader'' and moves to $q_2$. 

Rendez-vous communication and communication through a global store without a lock cannot produce a leader. Intuitively, the reason is that when
a process makes a move, arbitrarily many processes may follow suit, making 
exactly the same move immediately after. We will come back to this point later.

\section{The Power of Crowds}
\label{sec:power}

We can sort the four communication mechanism of the previous section
in order of decreasing power according to our two criteria:
\begin{center}
\begin{tabular}{ll}
broadcast communication & (everybody must listen, leader can be produced) \\
global store with locking & (somebody must listen, leader can be produced) \\
rendez-vous communication & (somebody must listen, no leader can be produced)\\
global store without locking & (nobody must listen, no leader can be produced)
\end{tabular}
\end{center}

In this section we show that this informal classification is confirmed by the mathematical results: the complexity of the coverability problem
decreases as we move down through the list. 

Before describing the results, it is important to observe that the complexity of the coverability problem is related to the crowd's computational power seen as a nondeterministic machine. If coverability is 
hard for a complexity class ${\cal C}$, then any problem in ${\cal C}$ can be reduced to coverability. Therefore, given an instance of the problem, we can construct a template $A$ such that a large enough crowd will solve it: a process will reach the state $q_u$, which now instead of an unsafe state becomes the state at which the process can post the answer ``yes''.  
So---informally but suggestively---studying the complexity of the coverability problem amounts to studying the following question:

\begin{quote}
{\bf What is the computational power of a (finite but arbitrarily large) anonymous crowd?}
\end{quote}

In particular, a result proving high complexity of the coverability problem 
means bad news for crowd verifiers, but good news for crowd designers,
and vice versa.

We are now ready to analyze  the complexity of the four communication 
mechanisms above.

\subsection{Communication by broadcast.}  Despite the power of 
broadcast communication, it was proved in \cite{EFM99} by Finkel, Mayr, and the author
that the coverability problem is decidable. So we have:

\vspace{0.3cm}
\begin{center}
\framebox{
\parbox{11.3cm}{
Anonymous crowds are not Turing powerful, or, conversely, identities are \\
necessary in order to achieve full Turing power.}}
\end{center}

The proof is a straightforward application of a more 
general result of \cite{abdulla96} on well-structured systems (see also \cite{abdulla00,finkel01}). Let us sketch it. The configuration of a crowd with
template $A$ is completely determined by the number of processes at each state of $A$. So, given a numbering $\{q_1,  \ldots, q_n\}$ of the
states of $A$, a configuration can be formalized as a vector of $\Nat^n$. Assume without loss of generality that 
$q_u=q_1$. We wish to know whether, for some number $N$, a crowd of $N$ individuals can reach a
configuration $(k_1, \ldots, k_n)$ such that $k_1 \geq 1$, or, equivalently, a configuration
$(k_1, \ldots, k_n) \geq (1,0,\ldots, 0)$, where $\geq$ is defined componentwise. 
The set of configurations $(k_1, \ldots, k_n) \geq (1,0,\ldots, 0)$ is {\em upward closed} (with respect to $\leq$), i.e., if a  configuration $c$ belongs to the set, then so does any other configuration of the form $c+c'$, where $c' \in \Nat^n$ and $+$ is defined componentwise.

Given an upward-closed set $C$ of configurations, it is easy to show that its set of immediate predecessors (i.e., the set of configurations from which some configuration of $C$ can be reached in one step) is also upward-closed.
Indeed, assume we can reach a configuration $c \in C$ from some configuration $d$ by means of the broadcast of a value $v$. Now, consider a configuration $d+d'$. If we perform the same broadcast, then the processes of $d$ move
to the same states as before, yielding again the configuration $c$, and the processes of $d'$ move somewhere, yielding a configuration
$c'$. The result is a configuration $c+c'$, where addition of configurations is defined componentwise. Since  $c \in C$ and $C$ is upward-closed, we have $c+c' \in C$, and we are done. So letting $C_0$ be the set of configurations $(k_1, \ldots, k_n) \geq (1,0,\ldots, 0)$, the sequence
$C_0, C_1, C_2, \ldots$, where $C_{i+1}$ is the set of immediate predecessors of $C_i$, is a sequence of upward-closed sets.

We now exploit the well-known fact that the order $\geq$ is a {\em well-quasi-order}: every infinite sequence $v_1, v_2, \ldots$ of elements of $\Nat^n$ contains an infinite ordered subsequence $v_{i_1} \leq v_{i_2} \leq \ldots$. 
A first easy consequence of the theory of well-quasi-orders is that any upward-closed set of configurations has finitely many minimal elements with respect to $\leq$. So, since an upward-closed set is completely determined by its minimal elements, we can use the  minimal elements as a finite representation of the set.  This allows to explicitly construct 
the sequence $C_0, C_1, C_2 \ldots$. A second easy consequence is that this sequence contains two indices $i < j$ such that $C_i \supseteq C_j$. So we can stop the construction at $C_j$, because subsequent steps will not discover any new configuration. The set $\bigcup_{k=0}^j C^k$ contains all configurations from which a configuration of $C_0$ can be reached. We can then inspect this set, and check whether it contains one of the possible initial configurations of a crowd.

So crowds communicating by broadcasts are not Turing powerful. 
But, how powerful are they? The answer, due to Schmitz and Schnoebelen 
\cite{DBLP:conf/concur/SchmitzS13}, is very surprising: 

\vspace{0.3cm}
\begin{center}
\framebox{
\parbox{11.3cm}{
The time complexity of the coverability problem for anonymous crowds communicating by broadcast
grows faster than any primitive recursive function.}}
\end{center}

More precisely, the result is that coverability of broadcast protocols is $\mathbf{F}_\omega$-hard, where $\mathbf{F}_\omega$ is a class of problems of ``Ackermannian complexity'' (i.e., whose complexity is bounded by an Ackermann-like function). In particular, $\mathbf{F}_\omega$ is closed under primitive recursive reductions. We refer to \cite{DBLP:conf/concur/SchmitzS13} for a more precise description. In any case, this is one of the most natural problem with provably non-primitive recursive complexity.

As a summary, we have that crowds communicating by broadcast 
may not be Turing powerful, but keeping them under control may quickly 
exceed any reasonable amount of computational resources.

\subsection{Communication by global store with locking.} 
Global variables with locking is the natural communication mechanism for multithreaded programs. 
The coverability problem for this kind of communication reduces to the coverability problem of 
Petri nets, a fact that was already observed by German and Sistla \cite{GS92},
and the converse also holds.

The coverability problem for Petri nets was proved to be EXPSPACE-complete
already in the 70s, which yields the following result:
\vspace{0.3cm}
\begin{center}
\framebox{
\parbox{11.3cm}{
The coverability problem for a crowd communicating by global variables with locking is  
EXPSPACE-complete.}}
\end{center}

EXPSPACE-hardness was proved by Lipton \cite{lipton76}
(see also \cite{DBLP:conf/ac/Esparza96}) who showed that a counter able to count up to $2^{2^n}$ can be simulated by a Petri net (or an automaton) of size $n^2$. Membership in EXPSPACE was proved by Rackoff \cite{DBLP:journals/tcs/Rackoff78}. He shows that, if the state $q_u$ is coverable, then it is coverable by a sequence of moves of double exponential length in the size of the template. This yields immediately an NEXPSPACE algorithm, after which we use NEXPSPACE=EXPSPACE.

Rackoff's nondeterministic algorithm is not useful in practice. A more practical algorithm was suggested (some years before Rackoff's paper) by Karp and Miller \cite{DBLP:journals/jcss/KarpM69}. The algorithm uses the notion of {\em generalized configuration}, which for a template with $n$ states is a vector of dimension $n$ whose elements are either natural numbers or the symbol $\omega$, which intuitively stands for ``arbitrarily many processes'', or ``as many process as necessary''. The algorithm starts at a generalized configuration describing the initial situation: for example, we may have exactly one process in state $q_1$, and arbitrarily many in state $q_2$, modelled by $(1,\omega,0, \ldots, 0)$. Given a generalized configuration, we construct its successors (that is, the algorithm explores new configurations in the forward direction, contrary to the algorithm for broadcasts, which explores backwards). If the template, say, has transitions $q_1  \by{v!} q_3$ and
$q_2 \by{v?} q_4$, then a rendez-vous can take place, and we can move from $(1,\omega,0, \ldots, 0)$ to
$(0,\omega,1,1,0,\ldots, 0)$. The important point is that this construction can be ``accelerated''. For example, if the template
has transitions $q_1  \by{v!} q_1$ and $q_2 \by{v?} q_4$, then we can move 
from $(1,\omega,0, \ldots, 0)$ to $(1,\omega,0,1,0,\ldots, 0)$ (state $q_2$ loses a process, but we apply
$\omega-1 = \omega+1=\omega$) and, since $(1,\omega,0,1,0,\ldots, 0) \geq (1,\omega,0, \ldots, 0)$, the rendez-vous can take place again, leading to $(1,\omega,0,2,0,\ldots, 0)$,  $(1,\omega,0,3,0,\ldots, 0)$,
etc. The algorithm ``jumps to the limit'', and moves directly from $(1,\omega,0, \ldots, 0)$ to $(1,\omega,0,\omega,0,\ldots, 0)$. Termination of the algorithm follows once more from a very simple application of the theory of well-quasi-orders.

Karp and Miller's algorithm has been recently improved in a number of ways: efficient data structures, construction of a minimal set of generalized configurations, etc. (see e.g. \cite{DBLP:conf/rp/PiipponenV13,DBLP:conf/apn/ValmariH12,DBLP:journals/ijfcs/GeeraertsRB10,DBLP:conf/apn/ReynierS11}). However, these improvements do not change its worst-case complexity, which is surprisingly worse than that of Rackoff's algorithm: Karp and Miller's algorithm can take non-primitive recursive time and space. Recently, this puzzling mismatch has lead to two beautiful results. First, Bozzelli and Ganty have shown that the backwards algorithm described above for broadcast systems no longer has non-primitive recursive complexity when applied to the rendez-vous case. Instead, it runs in double exponential time, much closer to the lower bound \cite{DBLP:conf/rp/BozzelliG11}. Geeraerts, Raskin, and Van Begin have proposed
another  simple algorithm based on forward exploration \cite{DBLP:conf/fsttcs/GeeraertsRB04}. It applies a so-called ``Enlarge, Expand, and Check'' algorithmic principle, which constructs a sequence of under- and overapproximations of the set of reachable generalized configurations.Very recently, Majumdar and Wang have shown that this algorithm also runs in 
double exponential time \cite{DBLP:conf/concur/MajumdarW13}.

Early work by Delzanno, Raskin and Van Begin \cite{DBLP:conf/tacas/DelzannoRB02} and more recent work by Kaiser, Kr\"oning and Wahl \cite{DBLP:conf/cav/KaiserKW10} (see also \cite{DBLP:journals/fmsd/DonaldsonKKTW12}) has applied these coverability algorithms and 
other techniques for the construction of over- and underapproximations, to verify safety of a large number of multithreaded programs.

\subsection{Communication by rendez-vous.} 
\label{subsec:rv}
Rendez-vous communication is a natural communication model for systems whose 
processes ``move'' in some medium where they occasionally meet and interact. 
Natural computing systems in which computing entities are molecules moving 
in a ``soup'' are an example. 

When studying the complexity of this problem there is a subtle point.
As we have seen in Section \ref{sec:paradigms}, a crowd communicating by rendez-vous communication 
cannot produce a leader. However, one can set up the system so that the 
initial configuration {\em already contains one}. For instance, we can choose 
an initial configuration with exactly one process in state $q_1$, and arbitrarily many processes in state
$q_2$. So we have to examine two cases.

\paragraph{Crowds with an initial leader.} In this case we can easily use rendez-vous to
simulate global store with locking. Intuitively, the template is designed so that
the leader simulates the store, and the rest of the crowd only communicates with the leader.
Conversely, as we saw in Section \ref{sec:paradigms}, rendez-vous 
communication can be simulated by a global store with locking, and so we obtain: 

\vspace{0.3cm}
\begin{center}
\framebox{
\parbox{11.3cm}{
The coverability problem for crowds communicating by rendez-vous and having an initial leader 
is EXPSPACE-complete.}}
\end{center}

\paragraph{Leaderless crowds.} This is the case in which 
all processes are initially in the same state. In other words, if we assume that this state is
 $q_1$, then the initial generalized configuration of the system is $(\omega, 0, \ldots, 0)$. 
We can again solve the coverability problem by means of the Karp-Miller algorithm.
However, it is easy to see that in this special case the algorithm 
can only generate new configurations whose components are either $\omega$ or $0$, and so
a configuration is completely determined by the set of $\omega$-components.
Moreover, a successor $(k_1', \ldots, k_n')$ of a generalized configuration $(k_1, \ldots, k_n)$ 
necessarily satisfies $k_i = \omega \Rightarrow k_i' = \omega$ for every $1 \leq i \leq n$, i.e.,
the set of $\omega$-components can only grow along a path.
Finally, the Karp-Miller graph satisfies a diamond property: if a configuration
with a set $\Omega_1$ of $\omega$-components has two successor configurations with different 
sets $\Omega_2$, $\Omega_3$ of $\omega$-components, then these two configurations have a 
common successor with set $\Omega_2 \cup \Omega_3$. These properties 
together yield a simple polynomial fixed point algorithm for the coverability problem: starting with
$\Omega = \{q_1\}$, let $\Delta(\Omega)$ be the set of states reachable from markings that 
put arbitrarily many tokens in every state of $\Omega$, and repeatedly execute 
$\Omega := \Omega \cup \Delta(\Omega)$ until a fixed point is reached. 

\vspace{0.3cm}
\begin{center}
\framebox{
\parbox{11.3cm}{
The coverability problem for leaderless crowds communicating by rendez-vous is in PTIME.}}
\end{center}

This case is studied in detail be German and Sistla in section 4 of \cite{GS92}, where they
show that many other analysis problems, not only coverability, can be solved in polynomial time.

\subsection{Communication by global store without locking.} 
\label{subsec:gswol}
Locking mechanisms are 
easy to implement in a multithreading environment where all threads are executed on a single processor, 
or on a number of processors physically close to each other. They become more problematic for crowdsourcing systems, ad-hoc networks, 
vehicular networks or, more generally, any sort of decentralized system where processes may enter or leave the system at any time. 
The danger of this setting is obvious: a process may acquire the lock, and leave the system without returning it, blocking the complete crowd. 
Additionally, the locking mechanism is not as
easy to implement as in a multithreading environment. 

The case of communication by global variables without locking 
has been recently investigated in \cite{DBLP:conf/cav/EsparzaGM13}. The main finding is that 
the absence of locking drastically simplifies the task of checking coverability (good news for verifiers), 
or, equivalently, decreases the computational power (bad news for designers):

\vspace{0.3cm}
\begin{center}
\framebox{
\parbox{11.3cm}{
The coverability problem for a crowd with initial leaders communicating by global 
variables without locking is NP-complete.}}
\end{center}

Intuitively, in the rendez-vous case the template can be designed so that 
a process communicates a value to, say, exactly three other processes, which 
allows the crowd to perform some arithmetic. In particular, the crowd can store an 
integer $n$ by putting exactly $n$ processes in a given state of the template. 
This is not possible in a global store without a lock, because the process has no 
control on how many processes may read a value.

The NP-completeness result is proved with the help of two lemmas. The first lemma 
shows that the crowd can be simulated by a system composed of a finite number of 
{\em simulators}, one for each value of $V$. The simulator for the value $v$ is an
automaton $A_v$ that can be easily constructed from the template $A$ and the value $v$.
So we can construct a finite crowd that simulates the behavior of any crowd 
with template $A$, of any size. This result already shows that the 
coverability problem is in PSPACE, but not yet that it belongs to NP.
Membership in NP is proved with the help of a second lemma. Loosely speaking, 
the lemma states that, if the unsafe state is reachable, then it can be reached 
by means of computations of the simulators that can be guessed in polynomial time. 

The leaderless case is, as in the case of rendez-vous, polynomial. Essentially, 
one uses the same algorithm.

\section{Some Results on Crowds of Infinite-State Processes.} 
\label{sec:extensions}
So far we have assumed that processes are finite state (i.e., the template is a finite automaton). 
If we totally relax this condition (for instance, if we allow processes to be Turing machines), 
then the coverability problem becomes of course undecidable: a crowd of one suffices to achieve Turing 
power! But we can consider milder extensions of the computational power of a process. 

For broadcast communication and global variables 
with locking, even very modest extensions already make the crowd Turing powerful.
In particular, this is already  the case if processes can count, i.e., if the template
is a finite automaton whose transitions may act on a counter, increasing or decreasing
it by one, or testing it for zero. Two processes suffice to simulate a two-counter machine,
which are known to be Turing powerful. A crowd can select a leader, who can then select a second leader,
and these two leaders can then communicate with each other, ignoring the messages from the
rest of the crowd. The same applies to rendez-vous if the crowd initially contains a leader. 

For global variables without locking, the situation is more interesting. 
In \cite{DBLP:conf/cav/EsparzaGM13} two extensions are considered. First, the paper studies the case 
in which processes are pushdown automata (since stack can be used as a counter, this includes the counter case). 
The coverability problems remains NP-complete for ``leaderless crowds'' and becomes PSPACE-complete for crowds with one leader. 

The second extension considers the case in which processes are Turing machines that 
can only run for polynomial time. This models the situation in which each process has
no restrictions in computational power, but can only contribute a polynomial amount 
of work to the crowd. Since the crowd is arbitrarily large, the total amount of work 
is not bounded, and so we could hope to be able to show problems far 
beyond NP. However, the coverability problem remains NP-complete. Interpreting
the result, we conclude that without a locking mechanism the crowd cannot distribute 
an arbitrary exponential computation among its members in such a way that each 
individual only does a polynomial amount of work. 

\paragraph*{Acknowledgements}

Very special thanks to Pierre Ganty, Jan K\v ret\'insk\'y, Michael Luttenberger, and
Rupak Majumdar for numerous comments on former versions of this note. In particular, 
Rupak suggested the final structure. 

Many thanks to Sasha Rubin for contacting me about the
errors in the former version, and for pointing out that the leaderless case for rendez-vous 
was studied in detail by German and Sistla in their seminal paper \cite{GS92}.



\bibliography{refsSTACS14}



\end{document}